\documentclass[conference,letterpaper]{IEEEtran}

\usepackage[utf8]{inputenc} 
\usepackage[T1]{fontenc}
\usepackage{url}
\usepackage{ifthen}
\usepackage{cite}
\usepackage{graphicx}
\usepackage{xcolor}
\usepackage{hyperref}
\usepackage[cmex10]{amsmath} % Use the [cmex10] option to ensure complicance
\usepackage{tabu}
\definecolor{mygreen}{rgb}{0.1,.6,0.1}
\usepackage{lipsum}% http://ctan.org/pkg/lipsum
\usepackage{multicol}% http://ctan.org/pkg/multicols

\usepackage{makecell}
\setcellgapes{2pt}
\usepackage[thinlines]{easytable}

\def\BibTeX{{\rm B\kern-.05em{\sc i\kern-.025em b}\kern-.08em
    T\kern-.1667em\lower.7ex\hbox{E}\kern-.125emX}}
\ifCLASSOPTIONcompsoc
    \usepackage[caption=false, font=normalsize, labelfont=sf, textfont=sf]{subfig}
\else
\usepackage[caption=false, font=footnotesize]{subfig}
\fi
\usepackage{lipsum}
\usepackage{tabu}
\providecommand{\keywords}[1]{\textbf{\textit{Index terms---}} #1}

\begin{document}
\title{Neural Normalized Min-Sum Message-Passing vs. Viterbi Decoding for the CCSDS Line Product Code} 

% %%% Single author, or several authors with same affiliation:
% \author{%
%   \IEEEauthorblockN{Stefan M.~Moser}
%   \IEEEauthorblockA{ETH Zürich\\
%                     ISI (D-ITET)\\
%                     CH-8092 Zürich, Switzerland\\
%                     Email: moser@isi.ee.ethz.ch}
% }

%%% Several authors with up to three affiliations:
\author{%
  \IEEEauthorblockN{Jonathan Nguyen*, Linfang Wang*, Chester Hulse*, Sahil Dani*, Amaael Antonini*, Todd Chauvin\ddag,\\ Divsalar Dariush\dag , Richard Wesel*}
  \IEEEauthorblockA{*Department of Electrical and Computer Engineering, University of California, Los Angeles, Los Angeles, California 90095\\
\dag Jet Propulsion Laboratory, California Institute of Technology, Pasadena, California 91109\\
\ddag SA Photonics, Los Gatos, California 95032\\
                    Email: \{nguyen.j, lfwang, chulse, sahildani, amaael, wesel\}@ucla.edu, \\ t.chauvin@saphotonics.com, Dariush.Divsalar@jpl.nasa.gov}
  %\and
  %\IEEEauthorblockN{Albus Dumbledore and Harry Potter}
  %\IEEEauthorblockA{Hogwarts School of Witchcraft and Wizardry\\
                    %Hogwarts Castle\\ 
                    %1714 Hogsmeade, Scotland\\
                    %Email: \{dumbledore, potter\}@hogwarts.edu}
}

%%% Many authors with many affiliations:
% \author{%
%   \IEEEauthorblockN{Albus Dumbledore\IEEEauthorrefmark{1},
%                     Olympe Maxime\IEEEauthorrefmark{2},
%                     Stefan M.~Moser\IEEEauthorrefmark{3}\IEEEauthorrefmark{4},
%                     and Harry Potter\IEEEauthorrefmark{1}}
%   \IEEEauthorblockA{\IEEEauthorrefmark{1}%
%                     Hogwarts School of Witchcraft and Wizardry,
%                     1714 Hogsmeade, Scotland,
%                     \{dumbledore, potter\}@hogwarts.edu}
%   \IEEEauthorblockA{\IEEEauthorrefmark{2}%
%                     Beauxbatons Academy of Magic,
%                     1290 Pyrénées, France,
%                     maxime@beauxbatons.edu}
%   \IEEEauthorblockA{\IEEEauthorrefmark{3}%
%                     ETH Zürich, ISI (D-ITET), ETH Zentrum, 
%                     CH-8092 Zürich, Switzerland,
%                     moser@isi.ee.ethz.ch}
%   \IEEEauthorblockA{\IEEEauthorrefmark{4}%
%                     National Chiao Tung University (NCTU), 
%                     Hsinchu, Taiwan,
%                     moser@isi.ee.ethz.ch}
% }

\maketitle

%%%%%%
%% Abstract: 
%% If your paper is eligible for the student paper award, please add
%% the comment "THIS PAPER IS ELIGIBLE FOR THE STUDENT PAPER
%% AWARD." as a first line in the abstract. 
%% For the final version of the accepted paper, please do not forget
%% to remove this comment!
%%
\begin{abstract}
The Consultative Committee for Space Data Systems (CCSDS) 141.11-O-1 Line Product Code (LPC) provides a rare opportunity to compare maximum-likelihood decoding and message passing. The LPC considered in this paper is intended to serve as the inner code in conjunction with a (255,239) Reed Solomon (RS) code whose symbols are bytes of data. This paper represents the 141.11-O-1 LPC as a bipartite graph and uses that graph to formulate both maximum likelihood (ML) and message passing algorithms.  ML decoding must, of course, have the best frame error rate (FER) performance. However, a fixed point implementation of a Neural-Normalized MinSum (N-NMS) message passing decoder closely approaches ML performance with a significantly lower complexity.

\medskip \noindent
\keywords{line product code, LDPC decoders, neural network, maximum likelihood, FPGA}
\end{abstract}

{\let\thefootnote\relax\footnote{{This research is supported by SA Photonics and the Air Force Research Lab (AFRL). Any opinions, findings, and conclusions or recommendations expressed in this material are those of the author(s) and do not necessarily reflect views of SA or AFRL. Research was carried out in part at the Jet Propulsion Laboratory, California Institute of Technology, under a contract with NASA.@2021. All rights reserved.}}}

\section{Introduction}
Line codes describe a set of encoding maps used to transmit digital data. The primary purpose of a line code is to manage the disparity of a transmission, which is defined as the difference between the number of transmitted $1$s and $0$s. Managing bit disparity has the benefit of minimizing DC components in transmissions which cannot be reliably transmitted over most long-distance communication channels. In this paper, we reference the Consultative Committee for Space Data Systems' (CCSDS) 141.11-O-1 proposed line code, known as the Line Product Code (LPC) \cite{book2018optical}.

%---Jonathan Maximum Likelihood part---------
While often impractical since their complexity scales at a rate of $O(2^n)$, maximum likelihood (ML) decoders represent the best possible decoding performance. Previous work including \cite{Wolf1978} and \cite{alma9941854073606533} have proposed methods of reducing the complexity of ML decoding for linear block codes by representing them as a trellis and performing Viterbi decoding. While still on the order of $O(2^n)$, these methods drastically reduce number of required operations, enough so that for a short blocklength code such as the LPC, ML decoding is considered.

%-----Linfang Message Passing Alg. part------
Message passing algorithms, such as belief propagation or MinSum, are low-complexity iterative decoders for linear block codes. However, message passing algorithms are sub-optimal because they assume that the Tanner graph defined by the parity check matrix has no cycles. As a result, for short block length codes with short cycles, message passing decoders do not provide satisfying performance.

Recently, numerous works have focused on improving the performance of message passing decoders with the aid of neural networks \cite{Nachmani2016-bs,Lugosch2017-ed,Nachmani2017-qq,Nachmani2018-ra,Liang2018-lw,Wu2018-zr,Lugosch2018-gu,Lyu2018-nz,wang2021neural}. Nachmani \textit{et al.} and Lugosch \textit{et al.} in \cite{Nachmani2018-ra, Lugosch2017-ed,Nachmani2016-bs} proposed Neural Normalized MinSum (N-NMS) and Neural Offset MinSum (N-OMS) decoders to improve the performance of the NMS and OMS decoders. Unlike NMS and OMS, which use a constant multiplicative or offset weight, N-NMS and N-OMS assign distinct trainable weights to each edge in each iteration. Simulations in \cite{Nachmani2018-ra, Lugosch2017-ed,Nachmani2016-bs} show that N-NMS and N-OMS have the capability to drastically improve the decoding performance of NMS and OMS for short-blocklength codes.

The rest of the paper is organized as follows: Section \ref{sec: encoding} introduces the LPC, how it manages bit disparity, and how it is encoded. Section \ref{sec: decoding} describes how to represent the LPC as a bipartite graph and how to perform both maximum likelihood and message passing decoding on it. Section \ref{sec: hardware} details the process of implementing the considered decoders on a Field Programmable Logic Array (FPGA). Simulation results and corresponding discussion are shown in Section \ref{sec: Simulation} and Section \ref{sec: conclusion} concludes this paper.

\section{Line Product Code Encoding} \label{sec: encoding}
The LPC encoder operates on blocks of 25 bits denoted by \textit{LPCEncIn}[24:0]. In particular, the most significant bit \textit{LPCEncIn}[24] is called channel system data and denoted by $S$. The LPC encoder discards \textit{LPCEncIn}[23:16] (legacy implementation detail of the laser communication terminal encoding process), and maps \textit{LPCEncIn}[15:0] to the following  $4\times 4$ matrix:
\begin{center}
    \begin{tabular}{|c|c|c|c|c|}
        \hline
        $u(0,0)$ & $u(0,1)$ & $u(0,2)$ & $u(0,3)$  \\
        \hline
        $u$(1,0) & $u$(1,1) & $u$(1,2) & $u$(1,3)  \\
        \hline
        $u$(2,0) & $u$(2,1) & $u$(2,2) & $u$(2,3) \\
        \hline
        $u$(3,0) & $u$(3,1) & $u$(3,2) & $u$(3,3)  \\
        \hline
    \end{tabular}
\end{center}

The LPC encoder generates the codewords of 24 bits. The codewords along with $S$ forms a $5\times 5$ matrix:

\begin{center}
    \begin{tabular}{|c|c|c|c|c|}
        \hline
        $e^*(0,0)$ & $e^*(0,1)$ & $e^*(0,2)$ & $e^*(0,3)$ & $ph(0)$ \\
        \hline
        $e^*(1,0)$ & $e^*(1,1)$ & $e^*(1,2)$ & $e^*(1,3)$ & $ph(1)$ \\
        \hline
        $e^*(2,0)$ & $e^*(2,1)$ & $e^*(2,2)$ & $e^*(2,3)$ & $ph(2)$ \\
        \hline
        $e^*(3,0)$ & $e^*(3,1)$ & $e^*(3,2)$ & $e^*(3,3)$ & $ph(3)$ \\
        \hline
        $pv(0)$ & $pv(1)$ & $pv(2)$ & $pv(3)$ &  $S$ \\
        \hline
    \end{tabular}
\end{center}

The encoding of LPC consists of the following steps:

\begin{enumerate}
    \item Calculate $e(i,j)$ ($i,j=0,\dots,3$) using \textit{LPCEncIn}[15:0] via differential encoding. In particular, we refer to $\{e(i,j)| i=0,1, j=0,\dots,3\}$ as sub-block 1 and $\{e(i,j)| i=2,3, j=0,\dots,3\}$ as sub-block 2.
    \item Calculate the horizontal parity bits $ph(i)$ ($i=0,\dots, 3$) and vertical parity bits $pv(i)$ ($i=0,\dots, 3$).
    \item Apply bit-wise inversion of sub-block 1 and/or 2 in order to minimize difference between the number of transmitted ones and zeros, which is also referred as disparity. We denote $e^*(i,j)$ ($i,j=0,\dots,3$) as the sub-block bits after inversion process.
\end{enumerate}

The following section describes these three steps in detail.

%In this section, we discuss the theory behind the LPC code. At the encoder's end, the input can be seen as a 16-bit binary vector mapped to a 4x4 matrix. However, this 4x4 matrix must be encoded based on a certain set of rules before passing it through a communication channel. We perform differential encoding on the 4x4 input matrix to obtain 4 horizontal and 4 vertical parity bits. When we add the horizontal and vertical parity bits and one bit of \emph{SysChnDat} to the 4x4 input matrix, we obtain a 5x5 encoded matrix block. The unencoded input matrix is shown below:

%In the matrix shown above, the u(4,0)\dots u(4,3) and u(0,4)\dots u(3,4) bits are discarded by the LPC Encoder. The S represents the \emph{SysChnDat} bit.

\subsection{Differential Encoding} 

Differential encoding is performed on the two sub-blocks separately. For sub-block 1, initialize $e(0,0) = u(0,0)$ and $e(1,0) = u(1,0) \oplus e(0,3)$:

\begin{equation}
    e(i,j) = u(i,j) \oplus e(i,j-1), \quad j > 0
    \label{equ: dif_equ}
\end{equation}

\begin{figure}
	\centering
	\includegraphics[width=0.6\linewidth]{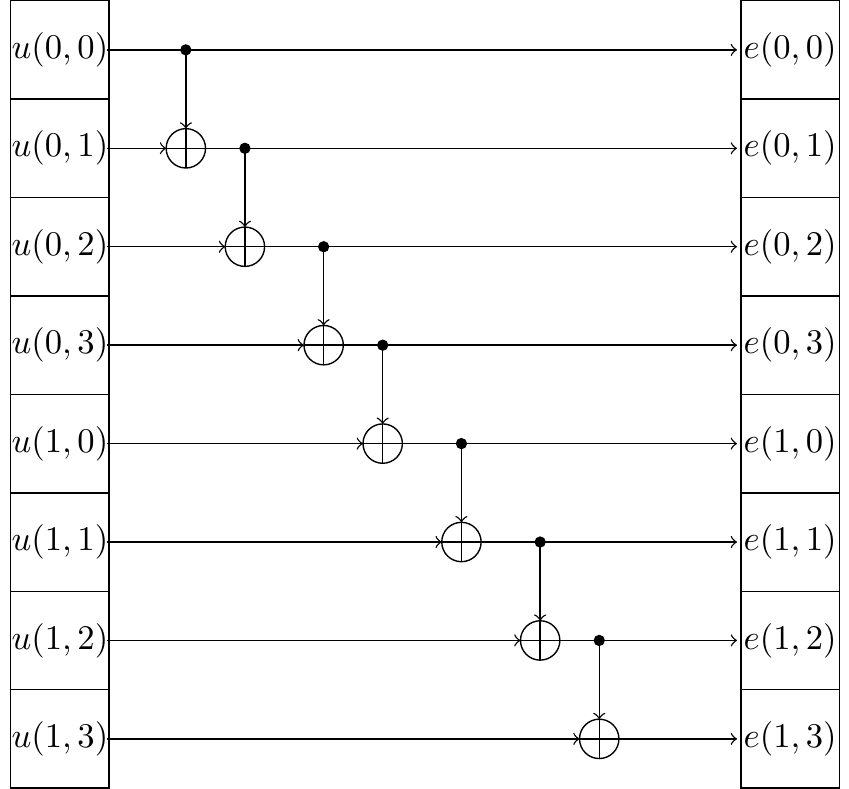}
	\caption{Differential encoding to calculate sub-block 1.}
	\label{fig: differential_encoding}
	
	\small
	\begin{tabular}{l}
	     where $\oplus$ = logical XOR
	\end{tabular}
\end{figure}

For sub-block 2, initialize $e(2,0) = u(2,0)$ $e(2,0) = u(2,0) \oplus e(0,3)$ %\eqref{equ: dif_equ}
as shown in  Fig. \ref{fig: differential_encoding}. The remaining bits can be derived using equation \eqref{equ: dif_equ}.  Fig. \ref{fig: differential_encoding} shows the differential encoding on sub-block 1.

\begin{figure*} 
\centering
  \includegraphics[scale = 0.4]{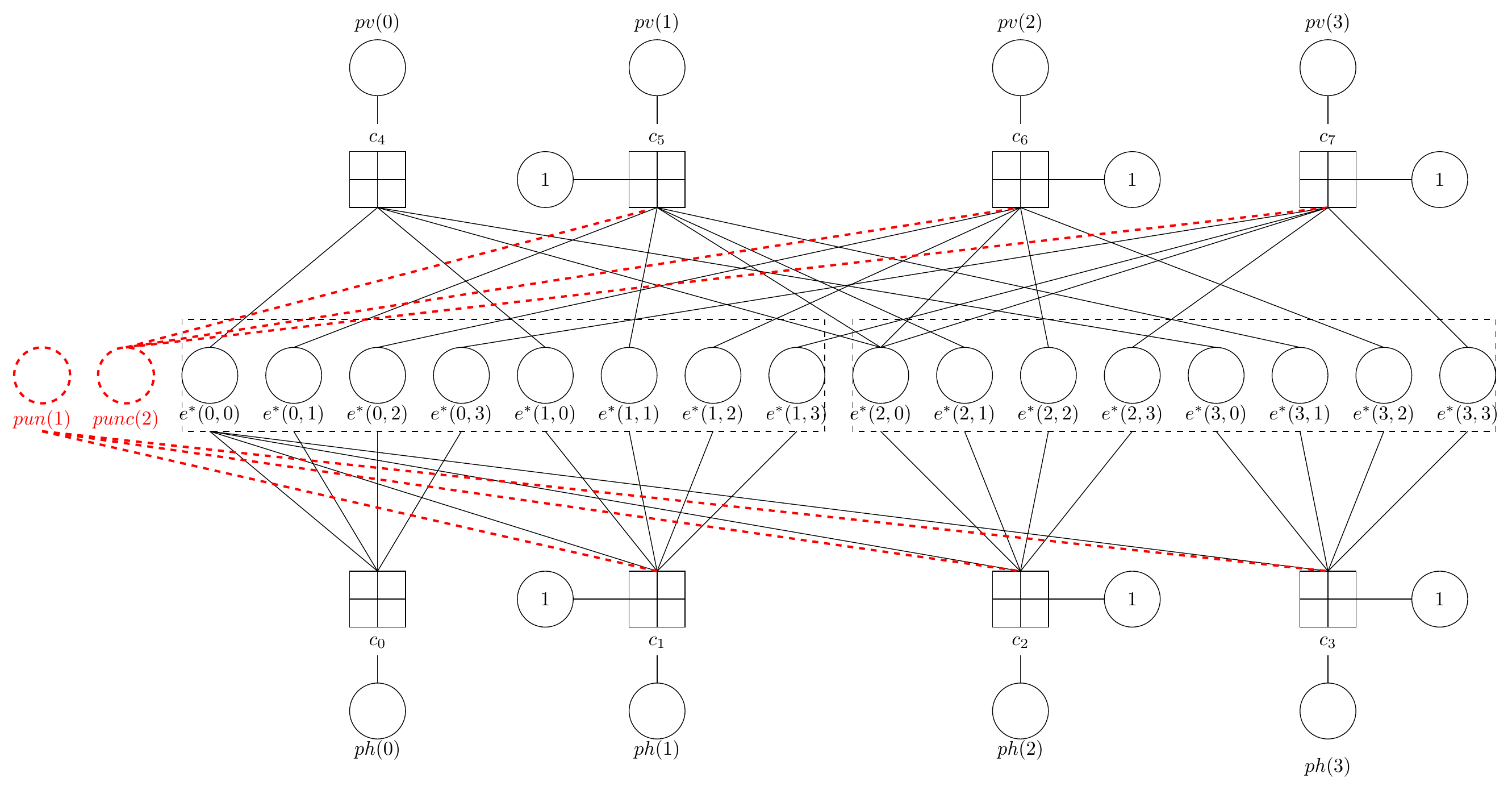}
  \caption{Bipartite graph of LPC. The red dashed circles are punctured variable node which indicate the sub-block inversion.}
  \label{fig: bipartite}
\end{figure*}

\subsection{Horizontal Parity bits}
After calculating the $\{e(0,0)\dots e(3,3)\}$ bits, the horizontal and vertical parity bits must be determined. The horizontal parity bit $ph(0)$ is always calculated for odd parity of the first row, meaning

\begin{align}\label{equ: hori-1}
    \bigg[ph(0)+\sum_{k=0}^{3} e(0,k)\bigg] \text{mod} 2 = 1
\end{align}

The other three parity bits $ph(1)$, $ph(2)$, and $ph(3)$ are determined not only by their corresponding rows, but also by $e(0,0)$. Specifically, if $e(0,0) = 0$, then odd parity is used for $ph(1)$, $ph(2)$, and $ph(3)$ and its corresponding rows. Otherwise, the even parity must be satisfied. Therefore, 

\begin{align}\label{equ: hori-2}
    \bigg[ph(i)+\sum_{k=0}^{3} e(i,k)\bigg] \text{mod} 2 = 1\oplus e(0,0) , ~ i=1,2,3.
\end{align}

% If e(0,0) = 0, then odd parity is used for ph(1), ph(2), and ph(3), meaning,
% $$\bigg[ph(1)+\sum_{k=0}^{3} e(1,k)\bigg] \% 2 = \bigg[ph(2)+\sum_{k=0}^{3} e(2,k)\bigg] \% 2 = $$ $$\bigg[ph(3)+\sum_{k=0}^{3} e(3,k)\bigg] \% 2 = 1$$
% If e(0,0) = 1, then even parity is used for ph(1), ph(2), and ph(3), meaning,
% $$\bigg[ph(1)+\sum_{k=0}^{3} e(1,k)\bigg] \% 2 = \bigg[ph(2)+\sum_{k=0}^{3} e(2,k)\bigg] \% 2 = $$ $$\bigg[ph(3)+\sum_{k=0}^{3} e(3,k)\bigg] \% 2 = 0$$

\subsection{Vertical Parity Bits }
The vertical parity bit $pv(0)$ is always calculated for even parity of the first row, meaning that

\begin{align}\label{equ: verti-1}
    \bigg[pv(0)+\sum_{k=0}^{3} e(k,0)\bigg] \text{mod} 2 = 0
\end{align}

The other vertical parity bits, $pv(1)$, $pv(2)$, and $pv(3)$ are calculated using their corresponding rows and $e(2,0)$. If $e(2,0) = 0$, then even parity is used for $pv(1)$, $pv(2)$, and $pv(3)$ and its corresponding rows. Otherwise, the odd parity must be satisfied. Therefore:

 \begin{align}\label{equ: verti-2}
     \bigg[pv(i)+\sum_{k=0}^{3} e(k,i)\bigg] \text{mod} 2 = e(2,0), ~i=1,2,3
 \end{align}
 
% $$\bigg[pv(1)+\sum_{k=0}^{3} e(k,1)\bigg] \% 2 = \bigg[pv(2)+\sum_{k=0}^{3} e(k,2)\bigg] \% 2 = $$ $$\bigg[pv(3)+\sum_{k=0}^{3} e(k,3)\bigg] \% 2 = 0$$
% If e(2,0) = 1, then odd parity is used for pv(1), pv(2), and pv(3), meaning,
% $$\bigg[pv(1)+\sum_{k=0}^{3} e(k,1)\bigg] \% 2 = \bigg[pv(2)+\sum_{k=0}^{3} e(k,2)\bigg] \% 2 = $$ $$\bigg[pv(3)+\sum_{k=0}^{3} e(k,3)\bigg] \% 2 = 1$$.

\subsection{Minimization of Disparity}
%Once the horizontal and vertical parity bits are calculated, we have all the information needed to construct the 5x5 output matrix. In the 5x5 matrix, Subblock 1 consists of nodes e(0,0)\dots e(1,3) and Subblock 2 consists of nodes e(2,0)\dots e(3,3). 
The disparity of each sub-block is defined as the difference between the number of transmitted ones and zeros. The goal of the LPC encoder is to minimize the disparity of each sub-block so that a relatively equal number of 0s and 1s are transmitted. Consequently, sub-block 1, sub-block 2, both, or neither are inverted at the encoder's end depending on the value of the disparity bits of each sub-block. Define \textit{DispSum}[$i$]($i=0,\dots,3$) as the disparity in the $5\times 5$ matrix, after the inversion of none, one or both sub-blocks. The following table lists inversion rules corresponding to each \textit{DispSum}[$i$] bit where $i=0,\dots,3$:

\begin{center}

\begin{tabular}{|c|c|c|}
\hline
                    & \begin{tabular}[c]{@{}c@{}}Inversion of \\ Sub-block 1\end{tabular} & \begin{tabular}[c]{@{}c@{}}Inversion of \\ Sub-block 2\end{tabular} \\ \hline
\textit{DispSum}[0] & No                                                                  & No                                                                  \\ \hline
\textit{DispSum}[1] & Yes                                                                 & No                                                                  \\ \hline
\textit{DispSum}[2] & No                                                                  & Yes                                                                 \\ \hline
\textit{DispSum}[3] & Yes                                                                 & Yes                                                                 \\ \hline
\end{tabular}

\end{center}

The LPC encoder performs sub-block inversion based on the rules shown in the previous table that provide minimum \textit{DispSum}[$i$]. Based on these inversion rules, the $e^*(i,j)$'s are calculated as follows:

\begin{align}
        e^*(i,j) &= 
     \left\{ \begin{array}{l l} 1-e(i,j)   & ~ \text{Inversion} \\  e(i,j)&  ~ \text{No Inversion} \\ \end{array}, \right.
\end{align}

where $i=0,1$ for sub-block 1, $i=2,3$ for sub-block 2, and $j=0,...,3$ for both sub-blocks.

% The calculation of \textit{DispSum}[0] \dots \textit{DispSum[3]} are referred to the Optical High Data Rate Communication Manual\cite{book2018optical}.
% NOTE TO SELF: FINISH CITATION HERE!!!!

\section{Line Product Code Decoding} \label{sec: decoding}
As a linear code, LPC can be represented by a parity check matrix $H$ and corresponding bipartite graph $\mathcal{G}$. Let $\mathbf{v}$ be a codeword of LPC, and define $\mathbf{s}(\mathbf{v})$ by: 
\begin{align}
    \mathbf{s}(\mathbf{v})= H\mathbf{v}^T.
\end{align}

Note that for a conventional linear block code, $\mathbf{s}(\mathbf{v})$ is a vector that is independent with $\mathbf{v}$. However, in this case, there are four distinct $\mathbf{s}$ with each one corresponding to a single inversion rule. One possible solution is to perform the decoding process using four different \textbf{$\mathbf{s}$} separately. This, however, will inevitably increase the hardware usage and decoding latency.

The following section shows that the four matrices can be combined into one by introducing two punctured variable nodes which indicate the inversion rule. As a result, decoding can be performed using only one matrix. The next sections describes the application of decoding methods including maximum likelihood and message passing to the LPC.

\subsection{Parity Check Matrix Representation}
Equations \eqref{equ: hori-1} through \eqref{equ: verti-2} put eight parity check constraints on $e(i,j)$, horizontal parity bits and vertical parity bits. The black, solid line portions in Fig. \ref{fig: bipartite} represent the bipartite graph defined by these eight parity checks. The "box-plus" symbols and circles represent check nodes and variable nodes, respectively. Circles with a 1 represent a special variable node whose value is a constant $1$. The eight check nodes are denoted as $c_0,...,c_7$. The bipartite graph is drawn such that the modulo-2 sum of all variable nodes connected to each check node must equal zero. These are known as the parity checks.

Given a valid codeword, incorrectly inverting one sub-block will cause the new codeword to fail some parity checks. More specifically, the check nodes that connect to an odd number of variable nodes in one sub-block will no longer satisfy all parity checks if that sub-block gets inverted. 

Therefore $\{c_1,c_2,c_3\}$ do not satisfy the parity check condition when sub-block 1 gets inverted and $\{c_5,c_6,c_7\}$ do not satisfy the parity check condition when sub-block 2 gets inverted. Two extra variable nodes \textit{punc}(1) and \textit{punc}(2) are introduced in order to make sure that the check nodes still satisfy the parity check condition after the sub-block inversion. \textit{punc}(1) connects the check nodes that have an odd number of variable node neighbors belonging to sub-block 1. When sub-block 1 gets inverted, \textit{punc}(1) equals 1 such that for each check node connected to \textit{punc}(1), all variable nodes connected to that check node sum to zero. Similarly, \textit{punc}(2) connects the check nodes that have an odd number of variable node neighbors belonging to sub-block 2. Fig. \ref{fig: bipartite} shows the complete bipartite graph of LPC.

\subsection{Maximum Likelihood Decoding via the Parity Check Matrix} \label{sec: ml_parity}
In this section, we describe how to utilize Wolf's work in \cite{Wolf1978}  to represent the Line Product Code (LPC) as a trellis for performing maximum likelihood (ML) decoding. Since ML decoding represents the theoretical limit of decoding performance, practical decoders which achieve frame error rates closer to it are more desirable. Furthermore, because the LPC is a short blocklength code, reduction in complexity via a trellis representation such as \cite{Wolf1978} may be feasible to implement in hardware.

Naive ML decoders simply compare the received codeword against all valid codewords. As a result, its complexity scales on the order of $2^k$ where $k$ is the number of information bits in the codeword. According to the bipartite graph for the LPC shown in Figure \ref{fig: bipartite}, there are $262,144$ unique codewords. For practical applications, this is infeasible with respect to hardware limitations, despite every calculation being independent and parallelizable.

As such, we leverage Wolf's framework in \cite{Wolf1978} to create a trellis representation of the LPC. While sacrificing some parallelism, the final trellis contains only $2,764$ edges which represents a 94-fold reduction in complexity compared to brute force ML decoding. To construct the trellis, we follow the procedure in \cite{Wolf1978} with one important caveat: we terminate the trellis at the syndrome of $(1,1,1,1,0,0,0,0)$ instead of the all-zeros syndrome. This is because the Line Product Code is NOT strictly linear due to the use of odd parity (XOR with a constant $1$). This means that the trellis does not terminate at the all-zeros syndrome, but rather at a syndrome with $1$s in the indices whose corresponding check node contains this constant 1, and $0$ elsewhere. In this case, the target syndrome is $(1,1,1,1,0,0,0,0)$. 

\begin{figure}[htb]
    \centering
    \includegraphics[width= 0.7\linewidth, trim = 20 20 20 20, clip]{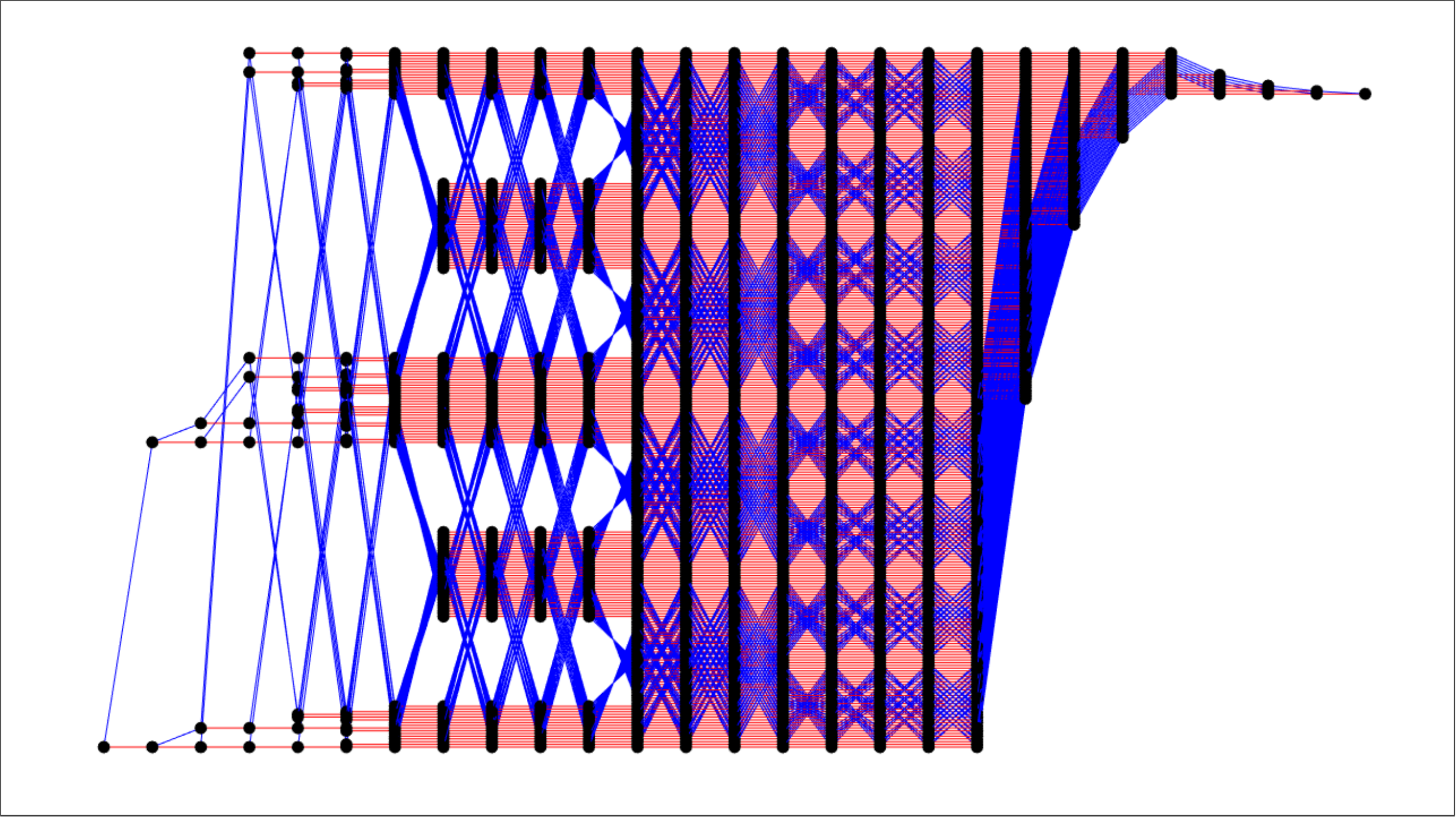}
        \caption{LPC Trellis Derived from Wolf's Method Following Pruning, with 2942 states and 5372 branches}
    \label{fig:lpc_trellis_trimmed}
\end{figure}

\subsection{Maximum Likelihood Decoding via the Generator Matrix} \label{sec: ml_generator}

The trellis representation of a code can also be constructed using a trellis oriented generator matrix \cite{alma9941854073606533}. A code's generator matrix $G$ can be transformed to become "trellis oriented" via row operations. This construction yields a minimal trellis for a given coordinate ordering. A permutation of the columns of $G$ yields a code $G'$ equivalent to $G$ on memoryless channels \cite{alma9941854073606533}. This permutation may yield a simpler trellis. However, finding this permutation is known to be an NP-hard problem. However, using heuristics, a permutation that simplifies the trellis may be found. As an example we show the graph of a trellis obtained with one such permutation of the Line Product Code. The new code yields a trellis with 1098 states and 1908 branches, down from 2942 states and 5372 branches of the original code.

% However, using heuristics, a permutation that simplifies the trellis may be found. As an example we show the graph of a trellis obtained with one such permutation of the Line Product Code, that yields a trellis with 1098 states and 1908 branches, down from 2942 states and 5372 branches of the original code.

% $\{1,  0,  3,  4,  5, 18,  2,  6, 11,  7, 22,  9,  8, 19, 14, 10, 15, 23, 20, 12, 13, 16, 24, 17, 21, 25\}$
% Original State profile:
% $\{1, 2,   4,   8,  6,  32,  64, 128, 128, 128, 128, 256, 256, 256, 256, 256, 256, 256, 256, 128,  64,  32,  16,   8,   4,   2,   1\}$
% Original Branch Profile:
% $\{2,	4,	8,	16,	32,	64,	128,	256,	256,	256,	256,	512,	512,	512,	512,	512,	512,	512,	256,	128,	64,	32,	16,	8,	4,	2\}$

% New State Profile:
% $\{1,   2,   4,   8,  16,  32,  32,  32,  64, 128, 128, 128, 128, 128,  64,  64,  32,  32,  16,  16,  16,   8,   8,   4,   4,   2,   1\}$
% New Branch Profile:
% $\{2,	4,	8,	16,	32,	64,	32,	64,	128,	256,	256,	256,	256,	128,	128,	64,	64,	32,	32,	32,	16,	16,	8,	8,	4,	2\}$
\begin{figure}[htb]
    \centering
    \includegraphics[width= 0.7\linewidth, trim = 20 20 20 20, clip]{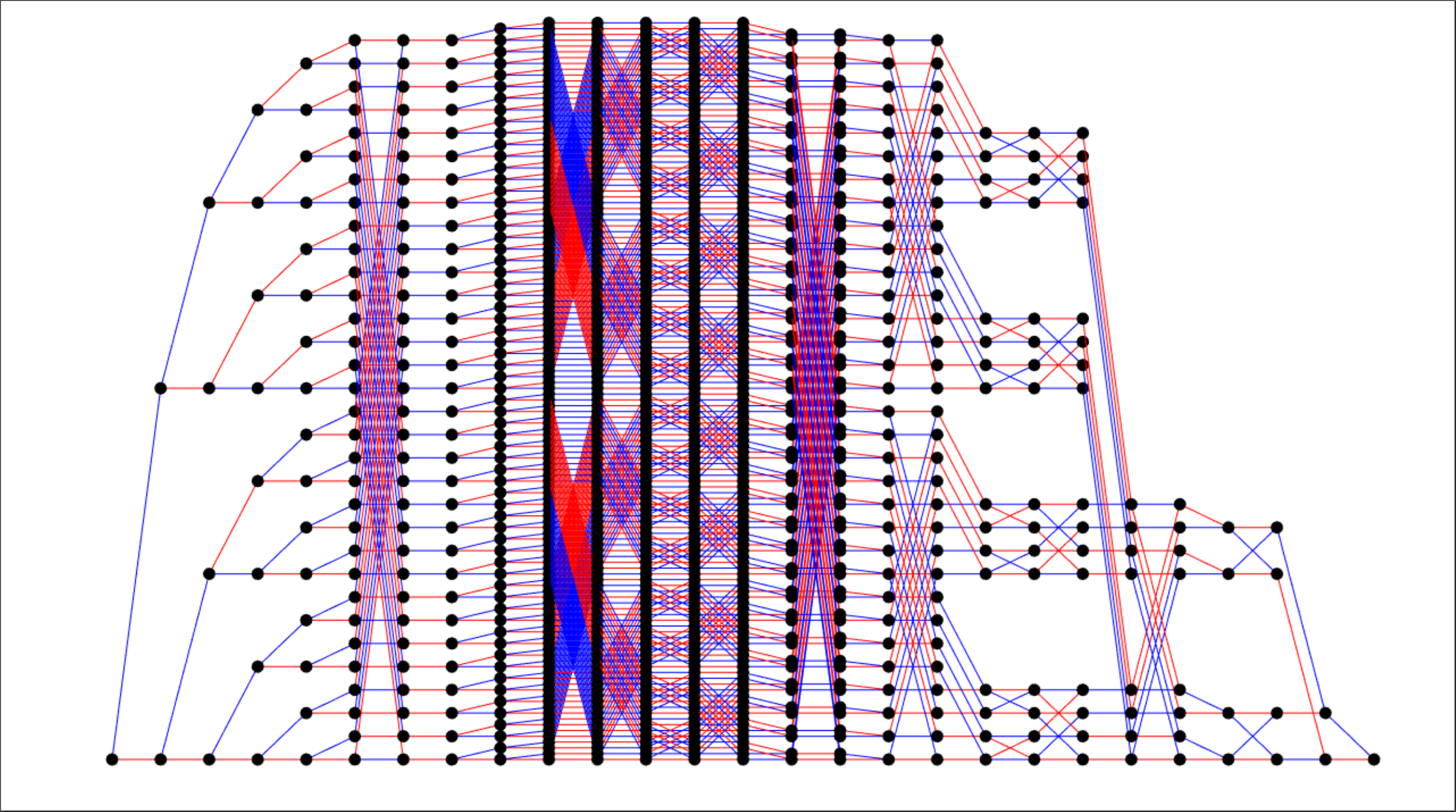}
        \caption{LPC Trellis Derived from a Generator Matrix Allowing Permutations, with 1098 states and 1908 branches}
    \label{fig:lpc_trellis_permutated}
\end{figure}

\subsection{Message Passing Decoding}
Message passing decoding algorithms, such as belief propagation and MinSum, provide an excellent decoding performance for linear block codes with large girths, defined as length of the shortest cycle in its Tanner graph. For the LPC, message passing decoders do not perform well, because its girth is only 4.

Recently, the neural-network-aided message passing decoders \cite{Nachmani2018-ra, Lugosch2017-ed,Nachmani2016-bs} have shown substantial improvements compared to conventional message passing decoders. Neural-network-aided message passing decoders assign distinct weights to each message in each iteration, such that the decoder can overcome trapping sets with short cycle lengths. This paper considers a neural normalized MinSum (N-NMS) decoder with a flooding schedule. In the $t^{th}$ decoding iteration, N-NMS updates the check-to-variable node message, $u^{(t)}_{c_j \rightarrow v_i}$, the variable-to-check node message, $l^{(t)}_{v_i\rightarrow c_j}$, and posterior of each variable node, $l_{v_i}^{(t)}$, by: 
\begin{align}
\begin{split}
       %u_{c_i\rightarrow v_j}&= \prod_{v_{j'}\in N(c_i)/\{v_j\}} \text{sgn}(l^{(t-1)}_{v_{j'}\rightarrow c_{i}}) \times \min_{v_{j'}\in N(c_i)/\{v_j\}} |(l^{(t-1)}_{v_{j'}\rightarrow c_{i}}|
    u^{(t)}_{c_i\rightarrow v_j} &= \beta^{(t)}_{(c_i,v_j)} \times  \prod_{v_{j'}\in \mathcal{N}(c_i)/\{v_j\}} \text{sgn}(l^{(t-1)}_{v_{j'}\rightarrow c_{i}}) \\ & \times  \min_{v_{j'}\in \mathcal{N} (c_i)/\{v_j\}} \left|(l^{(t-1)}_{v_{j'}\rightarrow c_{i}})\right| 
\end{split},
\\ 
l^{(t)}_{v_j\rightarrow c_i} &=  l^{ch}_{v_i} + \sum_{c_{i'}\in \mathcal{N}(v_j)/\{c_i\}} u^{(t)}_{c_{i'}\rightarrow v_j},\\
l^{(t)}_{v_j} &= l^{ch}_{v_i} + \sum_{c_{i'}\in \mathcal{N}(v_j)} u^{(t)}_{c_{i'}\rightarrow v_j}.
\end{align}

$\mathcal{N}(c_i)$ represents the set of the variable nodes connected to $c_i$ and $\mathcal{N}(v_j)$ represents the set of the check nodes that are connected to $v_j$. $l^{ch}_{v_j}$ is the LLR of the channel observation of $v_j$. $\beta^{(t)}_{(c_i,v_j)}$ are multiplicative weights to be trained. The decoding process terminates when all parity checks are satisfied or when the maximum iteration count, $I_T$, is reached. In this paper, we follow the steps of \cite{wang2021neural} to train the neural network.  

\section{Hardware Implementation of Message Passing Decoders} \label{sec: hardware}
 Despite ML decoding being the most optimal, its computational complexity for both the parity check and generator matrix derived trellises is too high to meet timing constraints for practical hardware implementation. Table \ref{tab: complexity} shows the worst case number of operations for each decoding method considered here. An operation is defined as an addition or multiplication, in the case of belief propagation (BP), we consider  $\arctan$, $\exp$, and $\log$ as a single operation since they are typically implemented via a Lookup Table (LUT). Table \ref{tab: complexity} indicates that message passing algorithms, except belief propagation, utilize significantly fewer operations than ML decoders, indicating that they are the most feasible to implement on hardware. It should also be noted that the Table \ref{tab: complexity} assumes that the message passing decoders always run for 8 iterations. In reality, for higher $E_b / N_0$, the number of required iterations approaches $1$, making message passing even more attractive. As such, our focus for this section will be on the N-NMS decoder, with MS and NMS decoders for the purpose of comparison. The field-programmable gate array (FPGA) device used for hardware implementation is the Zynq ZCU106 MPSoC.

\begin{figure}[t]
    \centering
    \includegraphics[width= 0.55\linewidth]{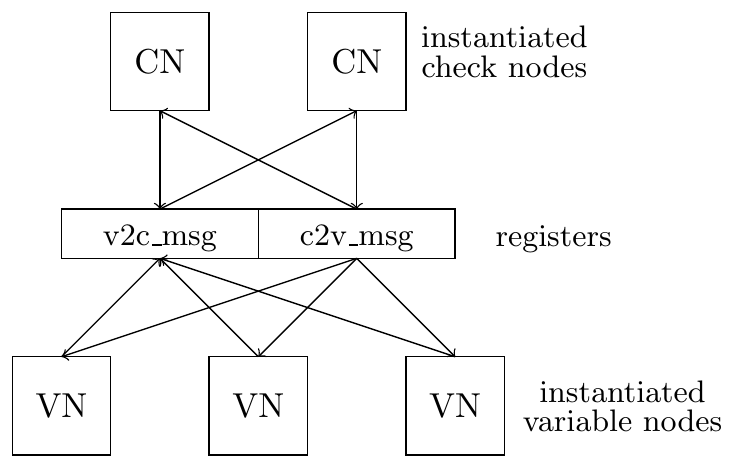}
    \caption{Block Diagram of FPGA architecture.}
    \label{fig:lpc_block_diagram}
\end{figure}

 The overall architecture consists of a bank of registers storing messages between check and variable nodes, and small instantiated modules to perform the respective check node (CN) and variable node (VN) operations, as seen in Figure \ref{fig:lpc_block_diagram}. The overall decoder controls the timing and coordinates the messages passed between the check and variable node modules for each decoding iteration. It also controls the terminating point by checking if the codeword estimate is valid or if the maximum number of iterations has been reached.

The initial FPGA implementation is a simple MinSum decoder, where check nodes search for the two minimums among its messages, and variable nodes compute simple summations. MinSum will be used as baseline to compare against the Normalized MinSum (NMS) and Neural-Normalized MinSum (N-NMS) implementations. 
% To modify this decoder into NMS we simply multiply the CN outputs by a constant. 
However, in order to implement the N-NMS decoder, we must dynamically assign edge weights depending on the iteration of the decoder. This task is divided between 2 modules: the main decoder and the check node module. The multiplicative edge weights are first quantized to a 6-2 scheme, meaning the first 6 bits represent the integer part of the number and the last 2 bits represent the fractional part. While other quantization schemes were considered, testing and simulations on the LPC showed that the 6-2 quantization achieved a satisfactory middle ground between accuracy and bit width.

Once the edge weights are quantized, they are stored in Block RAM (BRAM). The structure of the BRAM can represented as a 2-D matrix where each element represents a register that stores an 8-bit quantized edge weight. The index of a certain edge weight in the matrix also contains information regarding the iteration count and edge number in the bipartite graph. The main decoder module uses this information to assign weights to the proper edge depending on the iteration count. After the check node module calculates the check-to-variable node message, it then multiplies that by the incoming edge weight. This process repeats for every iteration until either a valid codeword is found or the maximum iteration count is reached. If a valid codeword cannot be attained by the maximum iteration, that codeword is declared un-decodable and the decoder gives up.

\begin{table}[t]
\caption{Comparison of Decoding Complexity via Number of Operations. }
    \begin{center}
        \begin{tabu}{|c|c|}
        \hline
            \textbf{Decoder} & {\scriptsize \textbf{Worst case number of operations}} \\
            \hline
            {\scriptsize BP*} & {10,192} \\
            \hline
            {\scriptsize standard MS*} & {3,200} \\
            \hline
            {\scriptsize NMS* / N-NMS*} & {3,616}  \\
            \hline
            {\scriptsize ML (Brute force)} & {524,288} \\
            \hline
            {\scriptsize ML (Parity check matrix trellis)} & {17,332} \\
            \hline
            {\scriptsize ML (Generator matrix trellis)} & {6,130} \\
            \hline
            \multicolumn{2}{c}{* Message passing algorithms are assumed to always run for 8 iterations}\\
        \end{tabu}
        \label{tab: complexity}
    \end{center}
\end{table}

\section{Simulation Results}\label{sec: Simulation}
\subsection{Frame Error Rates for Various Decoders}
In this section, we showcase our floating point simulation results for the Frame Error Rate (FER) of various decoding methods. The maximum likelihood FER was simulated using the trellis method described in sections \ref{sec: ml_parity} and \ref{sec: ml_generator}. Additionally, in line with practical limitations on actual decoding hardware, we limit the number of decoding iterations to two and eight.

The N-NMS decoder performed the closest to Maximum Likelihood out of the three decoders considered, even beating out Belief Propagation. These results line up with the findings of \cite{wang2021neural}. To summarize, via its training process, the N-NMS decoder was able to adapt its weights to the particular structure of the LPC unlike belief propagation or normalized MinSum. In particular, the N-NMS weights are specifically trained to mitigate the decoding loss caused by trapping sets. With the LPC being such a short block-length code, its cycles have particularly small girths making N-NMS the ideal decoding method for it.

\begin{figure}[htbp]
    \centering
    \includegraphics[width=0.8\linewidth] {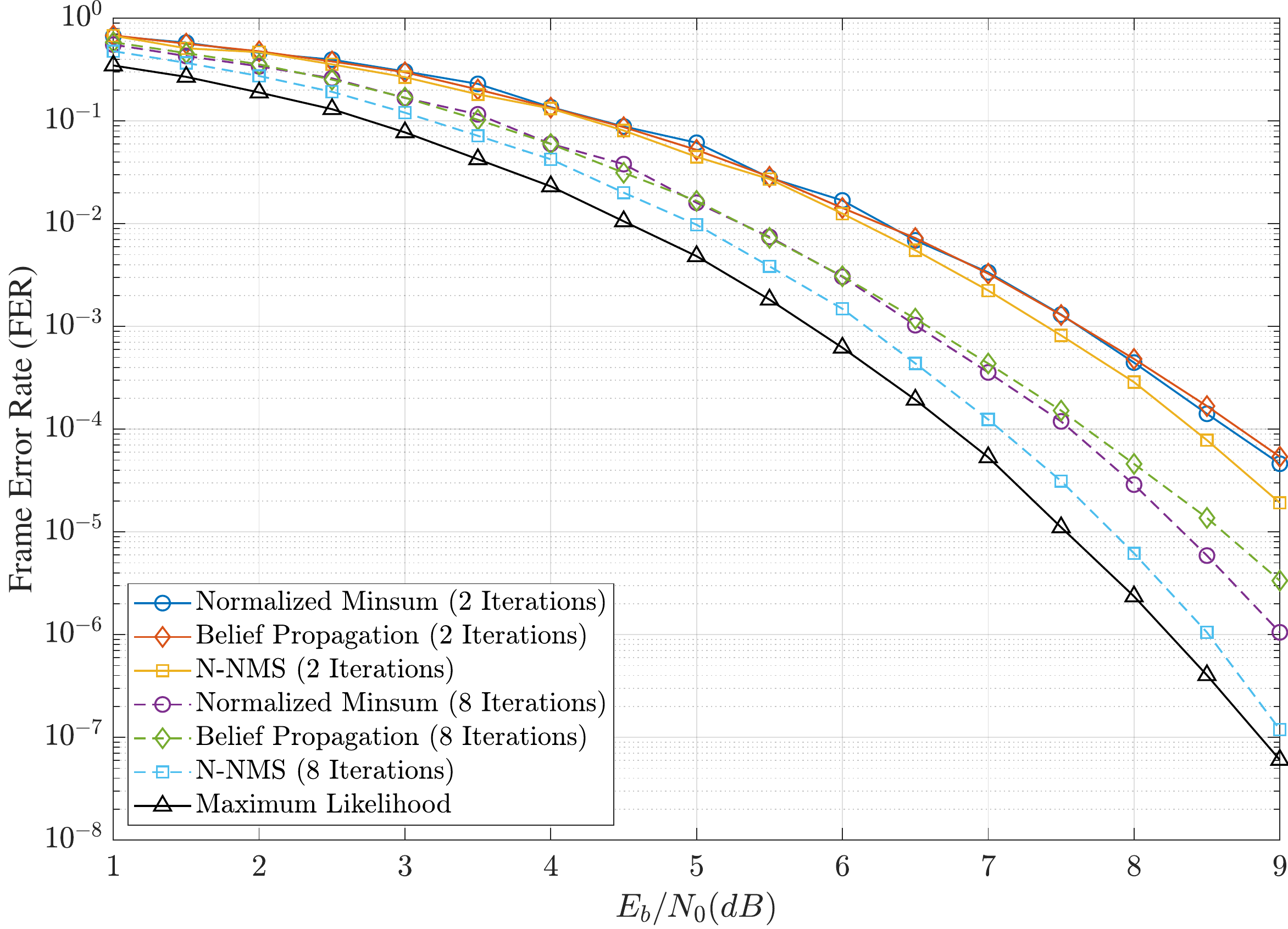}
    \caption{FER for various floating point decoding methods capped at 2 and 8 decoding iterations}
    \label{fig:fer_comparison_iter_2_8}
\end{figure}

\subsection{Quantization Loss for Fixed Point Decoders}
 In Table \ref{tab1}, we observe a noticeable increase in the Look Up Tables (LUTs) used by the N-NMS implementations as compared to the baseline MS and NMS ones. However, it is important to note that the N-NMS decoders also perform better than their counterparts. So, essentially, we are trading extra hardware utilization for better performance.
 
\begin{figure}[t]
    \centering
    \includegraphics[width=0.8\linewidth]{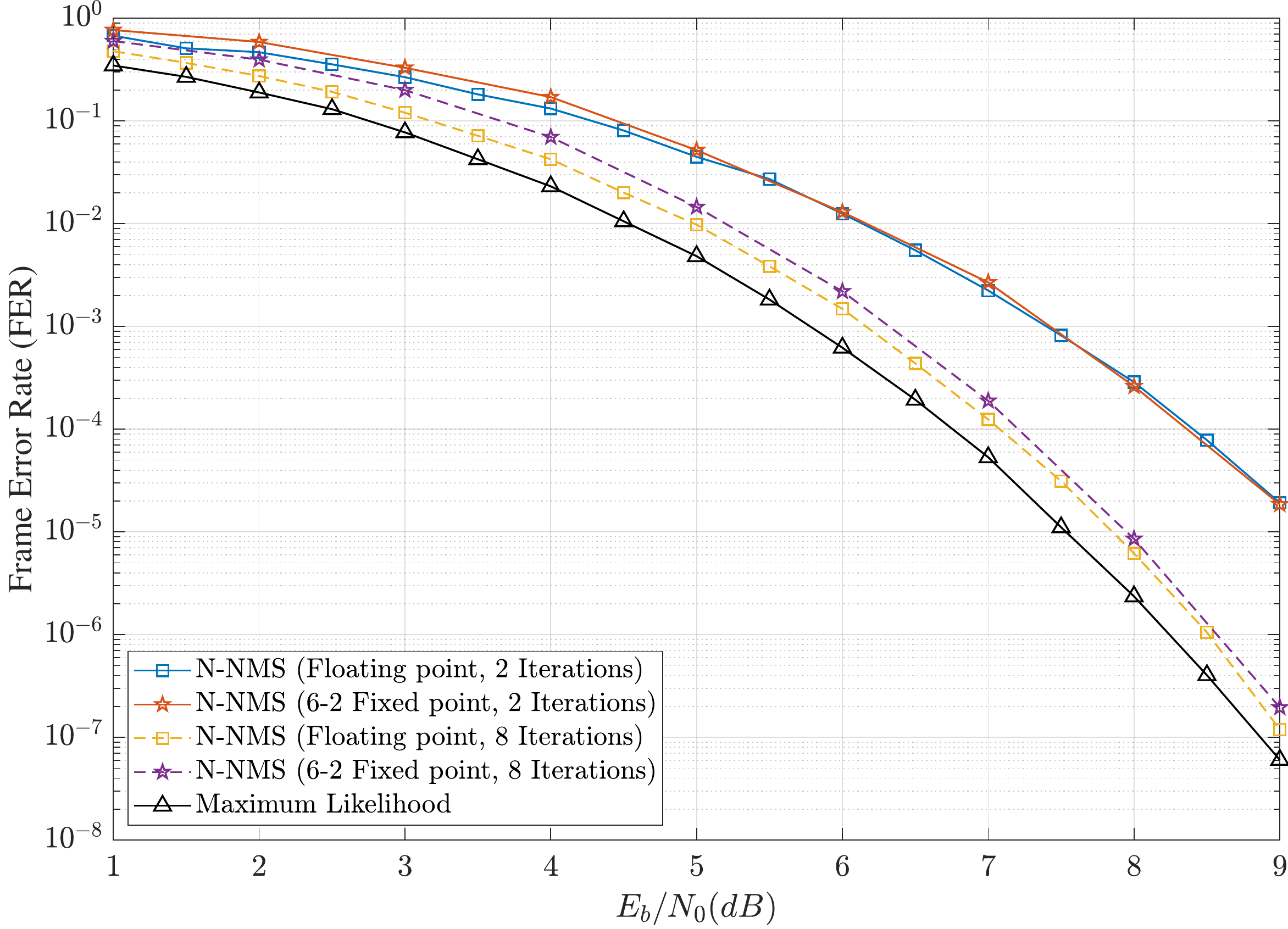}
    \caption{Floating vs fixed point FER for the N-NMS decoder capped at 2 and 8 decoding iterations}
    \label{fig:nnms_hardware_iter_2_8}
\end{figure}

\begin{table}[t]
\vspace{0.03in}
\caption{Decoder FER Performance and Resource Usage}
%\begin{center}
\begin{tabu}{|c|c|c|c|}
\hline
\textbf{decoder} & {\scriptsize \textbf{Eb/No (dB)$^{\mathrm{a}}$}} & \textbf{LUT} & \textbf{Reg.} \\
\tabucline[1.25pt]{-}
% still have to add all these number sin for real.
baseline MS (8) & 9.93 & 4953 (100\%) & 2201 (100\%)
\\ \tabucline[1.25pt]{-}
{\scriptsize NMS} & {9.63} & {5205 (105\%)} & {2201 (100\%)}  \\
\hline
{\scriptsize N-NMS(1$^{\mathrm{b}}$)} & {13.36} &  {5793 (117\%)} & {2201 (100\%)}  \\
\hline
{\scriptsize N-NMS(2)} & {10.54} &  {5784 (117\%)} & {2206 (100\%)} \\
\hline
{\scriptsize N-NMS(4)} & {9.28} & {5795 (117\%)} & { 2201 (100\%)}\\
\hline
{\scriptsize N-NMS(8)} &  {9.17} & {5796 (117\%)}  & { 2206 (100\%)}\\
\hline
\multicolumn{4}{c}{$^{\mathrm{a}}$Estimated $\left ( \frac{E_b}{N_o} \right ) $ to achieve FER of $10^{-7}.$ }\\
\multicolumn{4}{c}{$^{\mathrm{b}}$(n) number of iterations spent decoding.}
\end{tabu}
\label{tab1}
%\end{center}
\end{table}

The 6-2 quantization used on our FPGA is inherently different than typical software simulations which utilize 64-bit floating point numbers. Since we utilize fewer bits in our fixed point implementation, its precision is comparatively diminished to floating point and we expect some deterioration in FER. The purpose of the simulations shown in Figure \ref{fig:nnms_hardware_iter_2_8} is to demonstrate that with the N-NMS decoder, utilizing a fixed point quantization as opposed to floating point presents an almost negligible loss in frame error rate.

\subsection{Reed Solomon Frame Error Rate}
\begin{figure}[t]
    \centering
    \includegraphics[width = 0.8\linewidth]{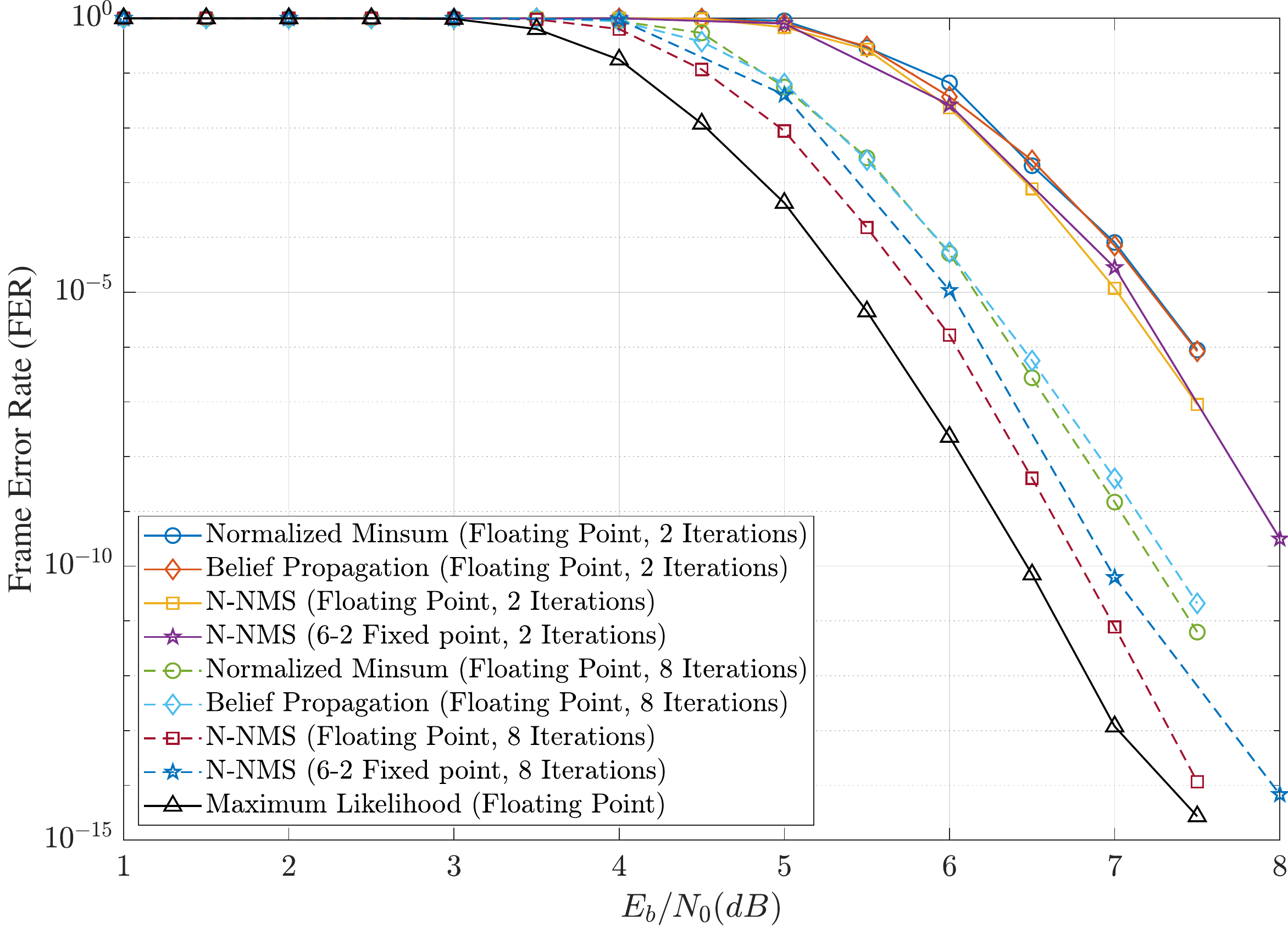}
    \caption{Reed Solomon Code FER for Floating vs Fixed point N-NMS and Maximum Likelihood Decoding on the LPC}
    \label{fig:rs_fer}
\end{figure}

As noted in the CCSDS specification, the LPC serves as the inner code in conjunction to a (255,239) Reed-Solomon (RS) operating on GF(256). Since the RS code has $16$ parity bytes, it can correct for up to $8$ byte errors. Considering that the LPC encodes $2$ bytes of data at a time, the RS code can correct for up to $4$ LPC frame errors. Therefore, given the FER of the LPC, the corresponding FER of the RS code can be modeled via a binomial expression: $P_{RS}(e) = 1 - P(X < 5), \text{ where } X \sim B(128, P_{LPC}(e))$. Figure \ref{fig:rs_fer} shows the Reed Solomon FER for the N-NMS fixed and floating point implementations at various $E_b/N_0$ with the LPC as the inner code. 

\section{Conclusion} \label{sec: conclusion}
This paper compares both the feasibility and decoding performance of maximum likelihood, belief propagation, standard MinSum, normalized MinSum, and Neural Normalized MinSum (NNMS) decoders on the Line Product Code (LPC). An initial exploration of maximum likelihood decoding was considered due to the LPC's short blocklength. However, attempted simulation on an FPGA board showed that, even with complexity reduction via a trellis, ML decoding failed to meet timing requirements. In lieu of this, we considered message passing algorithms, while less optimal than ML decoding, can be performed iteratively and with fewer operations than it. Simulation results on the LPC show that, with sufficient iterations, these message passing algorithms approach the frame error rate achieved by ML decoding. In particular, the NNMS decoder, with only 8 iterations, show only a 0.5 dB loss compared to ML making it the most promising decoder considered here.

\bibliographystyle{IEEEtran}
\bibliography{lpc_references,nn_decoder}

\end{document}